\documentclass{ws-procs9x6}
\input babarsym

\begin{document}

\title{MEASUREMENT OF $e^+e^-$ MULTIHADRONIC CROSS SECTIONS
BELOW $4.5$ GeV WITH \babar}

\author{Achim Denig\\
Representing the \babar~ collaboration}

\address{Institut f\"ur Experimentelle Kernphysik, Universit\"at Karlsruhe (TH)\\
Postfach 3640, D-76021 Karlsruhe, Germany\\
E-mail: achim.denig@iekp.fzk.de}
%
%
%
\begin{abstract}
We present a summary of the hadronic cross section measurements performed
with \babar~ at the PEP-II collider via {\it radiative return}. 
\babar~ has performed measurements of exclusive final states containing 3, 4 and 6 hadrons
via this complementary method, as well as a measurement of the proton form factor.
\end{abstract}


\bodymatter

\section{Initial State Radiation Physics at \babar}\label{aba:sec1}
At the particle factories DA$\Phi$NE and PEP-II 
the hadronic cross section  $\sigma(e^+e^-\to {\rm hadrons})$ is measured 
over a wide energy range by {\it radiative return} \cite{dafne}$^,$\cite{review} . In this new method only 
those events are considered, in which one of the beam electrons or positrons has
emitted an initial state radiation (ISR) photon, lowering in such a way 
the effective invariant mass of the hadronic system.  
Precision measurements of the hadronic cross section are of utmost importance since they provide
input to data-driven calculations of the hadronic contributions to the 
anomalous magnetic moment of the muon, $a_\mu$, and of the running fine
structure constant $\alpha(m_Z^2)$ \cite{ej}$^,$\cite{dehz} . In this paper we present measurements 
of different exclusive final hadronic states in the mass range $<4.5$ GeV, 
performed at the B-factory PEP-II ($\sqrt{s}=10.6$ GeV) with the detector 
\babar.
At \babar~ the ISR photon is required to be emitted at large polar angles with
respect to the beam axis,
allowing a kinematic
closure of the event ({\it tagging}). Since the hadronic system is recoiling opposite
to the ISR photon, a measurement of the cross sections 
with very high geometrical acceptance becomes possible.
In order to extract the {\it non-radiative} 
cross section from the measured {\it radiative} cross section, one 
normalizes to a well-known radiator function \cite{phokhara} and to the PEP-II
integrated luminosity, or - alternatively - to the yield of 
$e^+e^- \to \mu^+\mu^-\gamma$ events.

\begin{figure}
\hspace*{.2truecm}
  \includegraphics[height=.25\textheight]{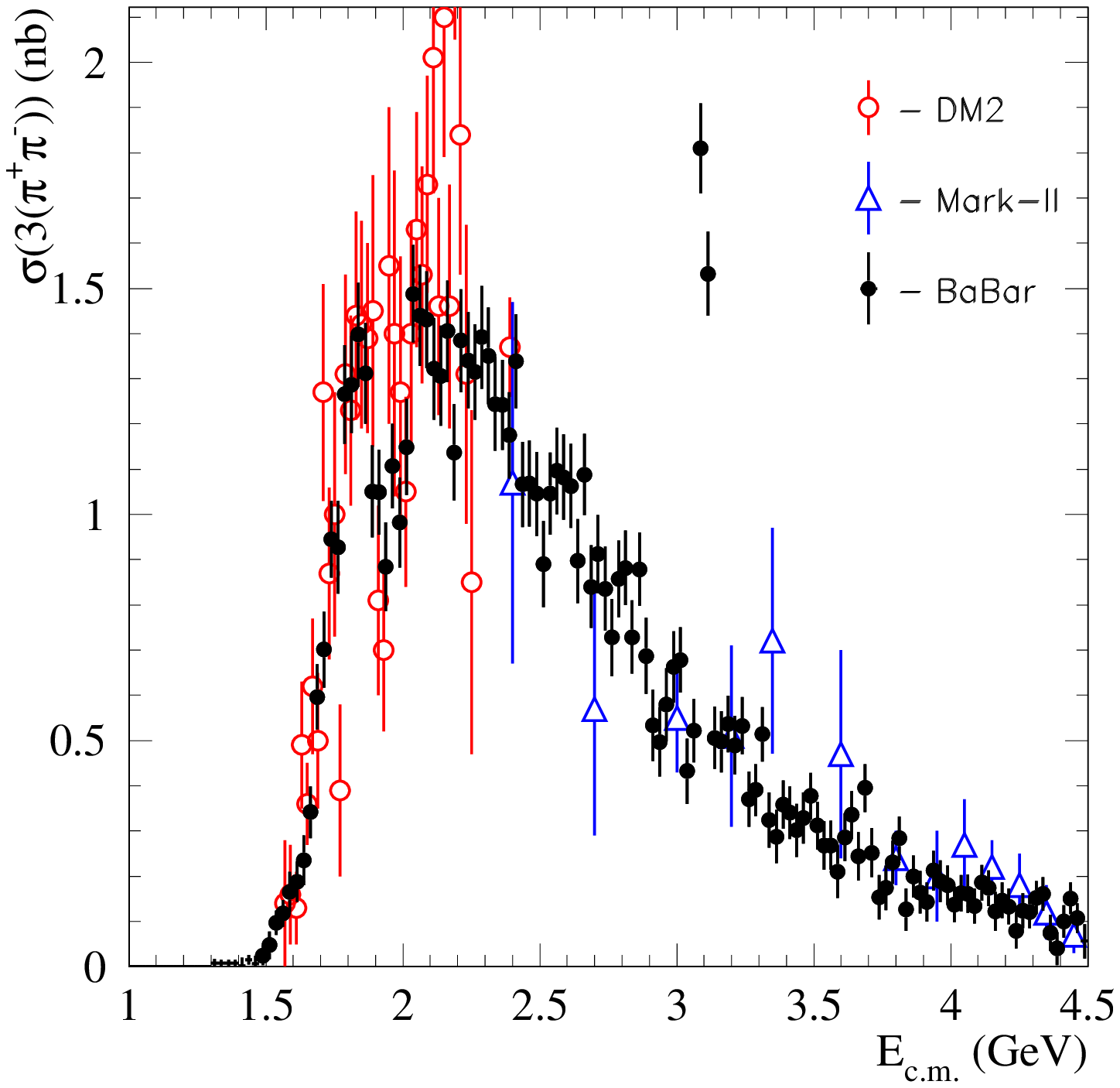}
\hspace*{1.truecm}
  \includegraphics[height=.25\textheight]{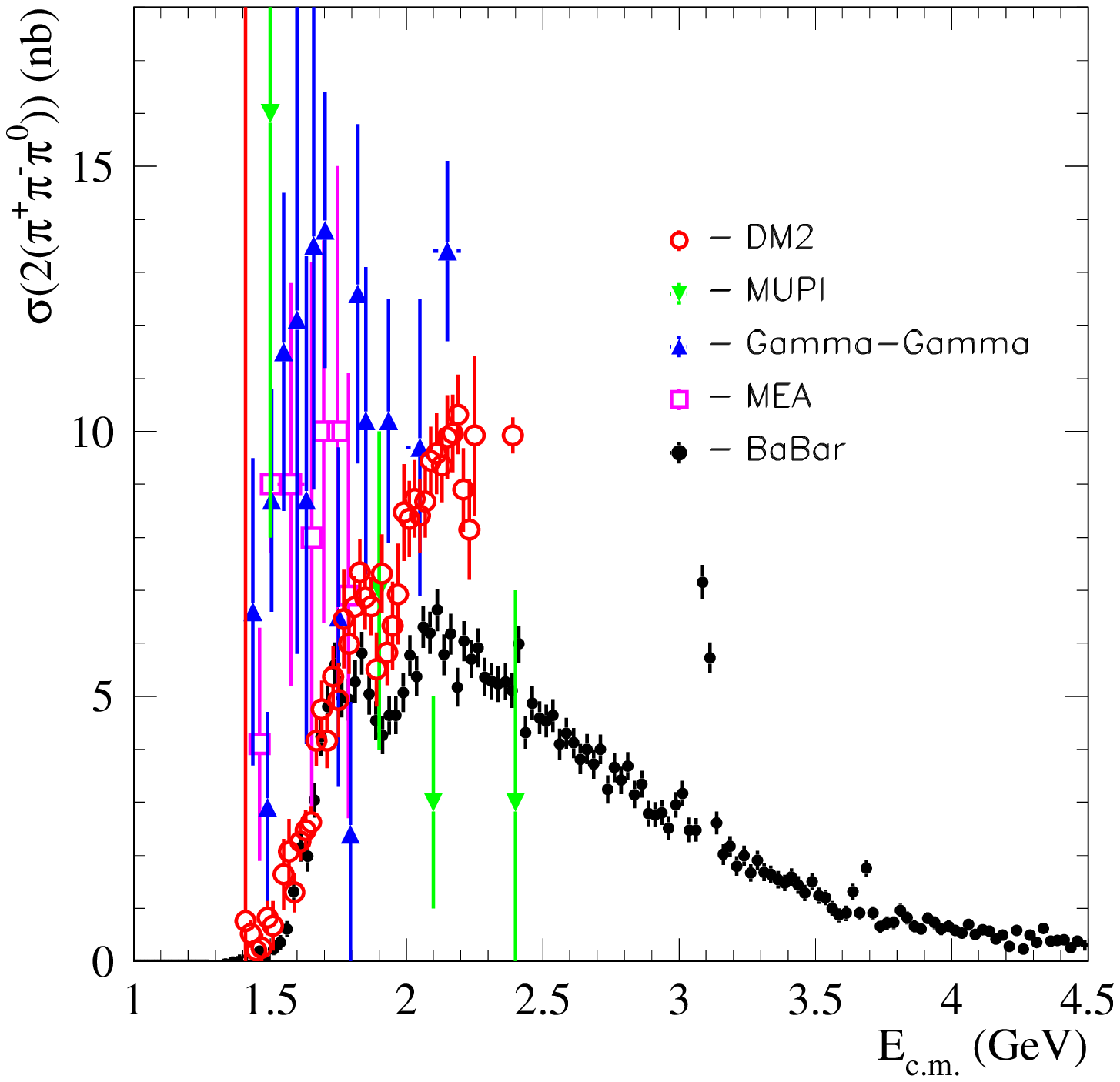}
  \caption{
  \label{6pixs}
The \babar~ measurement of the cross section for 
$e^+e^-\to \pi^+\pi^-\pi^+\pi^-\pi^+\pi^-$ (left) and 
$e^+e^-\to \pi^+\pi^-\pi^+\pi^-\pi^0\pi^0$ (right) vs. $\sqrt{s}$
compared to the world data set.  
}
\end{figure}

\section{Results}

{\bf Three and four hadrons}\\
\babar~has previously published measurements \cite{3pi}$^,$\cite{4pi} of the $\pi^+\pi^-\pi^0$,
$\pi^+\pi^-\pi^+\pi^-$, $K^+K^-\pi^+\pi^-$, $K^+K^-K^+K^-$ final states with
better precision and coverage than all previous experiments, using $89$ ${\rm fb}^{-1}$
of data. The systematic accuracy of the $3\pi$- and $4\pi$-channels in
the mass region $1-2$ GeV is $5\%$. All states have been studied also in terms
of their internal structures. In the $\pi^+\pi^-\pi^0$ analysis it was possible
to improve significantly on the world's knowledge the excited $\omega$ states, while
in the $\pi^+\pi^-\pi^+\pi^-$ channel a very strong contribution from the two-body
mode $a_1(1260)\pi$ was identified. Preliminary results from a data sample 
of $232$ ${\rm fb}^{-1}$ are available for the 
$e^+e^-\to K^+K^-\pi^+\pi^-$ and $K^+K^-\pi^0\pi^0$ cross section \cite{KKpipi} .
In the $\phi(1020)f_0(980)$ intermediate two-body state 
a vector resonance-like structure is 
seen near threshold with a mass of $(2175\pm10_{\rm stat}\pm15_{\rm syst})$ MeV and a width 
of $(58\pm16_{\rm stat}\pm20_{\rm syst})$ MeV.  
\\
\\
{\bf Six hadrons}\\
The six-hadron process \cite{6pi} has been studied in a data sample of $232$ ${\rm fb}^{-1}$ in the channels
$e^+e^-\to 3(\pi^+\pi^-), 2(\pi^+\pi^-\pi^0)$ and $2(\pi^+\pi^-)K^+K^-$. 
The cross sections for the first two channels are shown in fig.~\ref{6pixs};
large improvements over existing measurements are seen, as well as a much
wider coverage of the mass range. In the all-charged mode very little 
substructures have been found; a simulation containing one $\rho^0$ and
four pions distributed according to phase space is adequate to describe
the internal structure. On the contrary the partly neutral state shows a much
more complex structure with signals for $\rho^0, \rho^\pm, \omega$ and $\eta$, and
a substantial contribution from $\omega \eta$, which seems to be resonant.
In both channels a structure at ca. 1900 MeV, which had previously been seen 
by DM2 and FOCUS \cite{focus} , is clearly visible. Fits to the $3(\pi^+\pi^-)$ and 
$2(\pi^+\pi^-\pi^0)$ spectra, assuming a resonant structure over
a continuum shape, give consistent results for the mass $M$ and width $\Gamma$ of the structure.
For the channel $3(\pi^+\pi^-)$ we find $M=(1880\pm30)$ MeV and $\Gamma=(130\pm30)$ MeV,
for the channel $2(\pi^+\pi^-\pi^0)$ $M=(1860\pm20)$ MeV and $\Gamma=(160\pm20)$ MeV.
\\
\begin{figure}
  \includegraphics[height=.3\textheight]{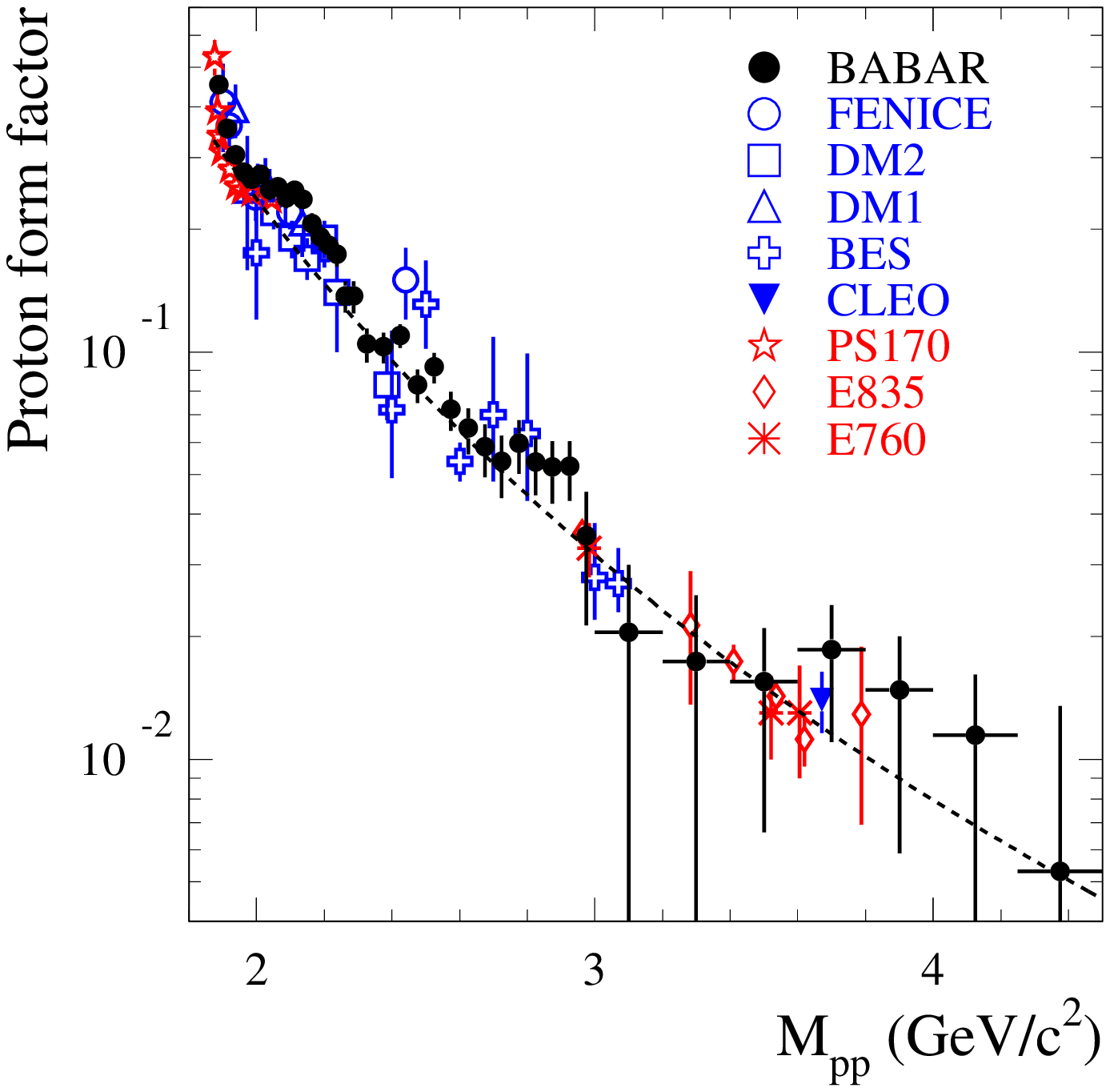}
  \includegraphics[height=.3\textheight]{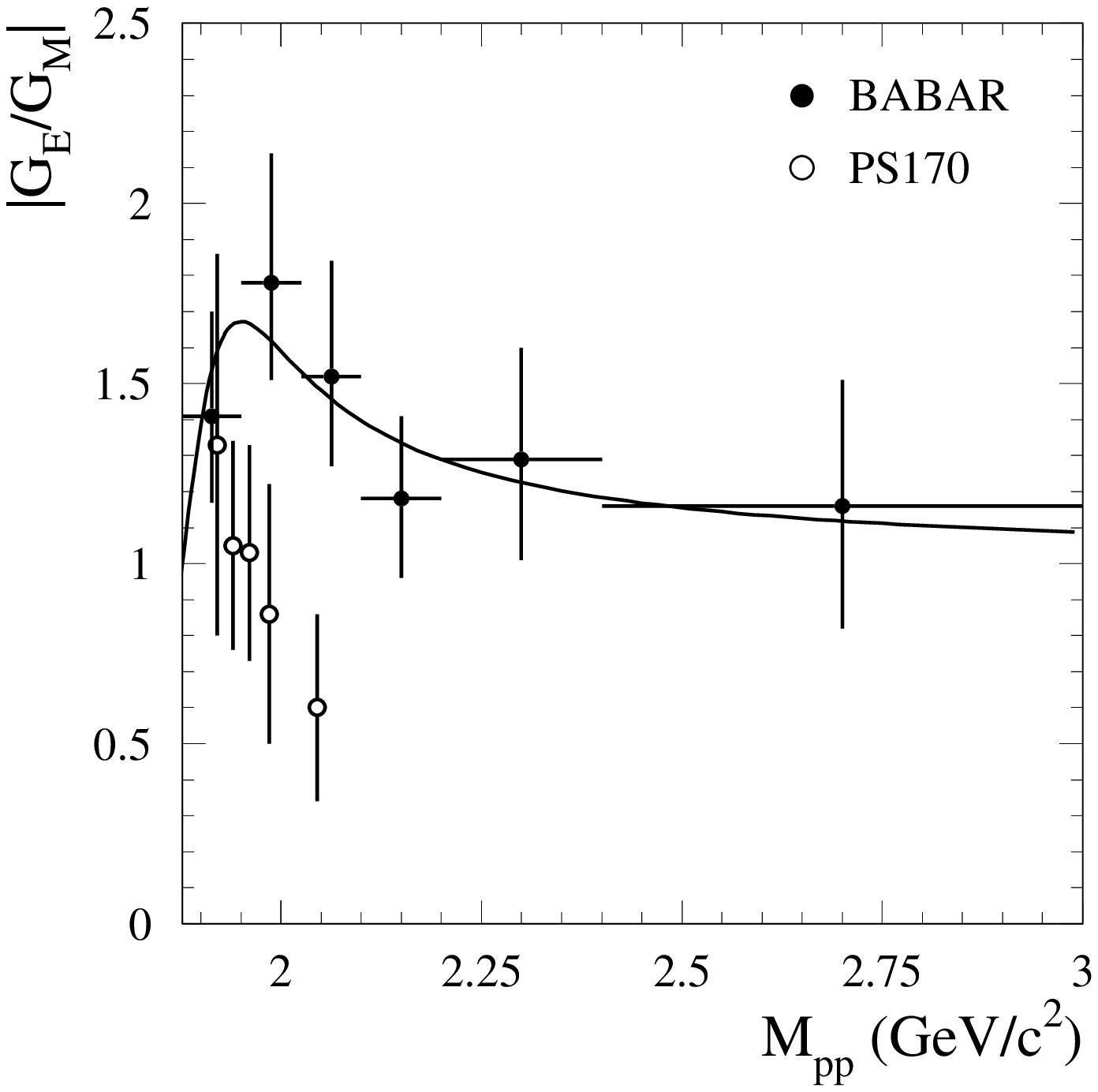}
  \caption{
  \label{ppbarff}
Left: The \babar~ measurement of the effective proton form factor  
and previous data obtained in $e^+e^-$ and $p\bar{p}$ experiments.
Right: The ratio of the electric and magnetic form factor $|G_E/G_M|$ as a function of 
$\sqrt{s}$.
}
\end{figure}
\\
{\bf Proton form factor}\\
\babar~ has also measured the cross section $e^+e^-\to p\bar{p}$ using
$240$ ${\rm fb}^{-1}$ of data \cite{pp} ; the corresponding effective form factor is shown
in fig.~\ref{ppbarff} (left), along with previous data from $e^+e^-$ and $p\bar{p}$ experiments.
We find an overall good consistency. The mass dependence shows a significant 
threshold enhancement, as well as two structures featuring sharp drops at 
$2.25$ and $3.0$ GeV, which illustrate the power of data from one single 
experiment over a wide range with no point-to-point uncertainties. 
Measuring the proton helicity angle $\theta_P$ in the $p\bar{p}$ rest frame,
one can separate the ratio of the electric and magnetic form factor 
$|G_E/G_M|$, since both show a different functional behaviour in $\theta_P$.
The \babar~ measurement of this ratio is shown in fig.~\ref{ppbarff} (right) 
for six different mass bins of $M_{p\bar{p}}$; a previous 
LEAR measurement \cite{lear} is in disagreement with \babar. 
Our data shows a significant increase of
the ratio $|G_E/G_M|$ towards threshold, while the two form factors 
are consistent at larger masses.

\section{Conclusions}
Measurements of the hadronic cross section at PEP-II 
have improved the knowledge of the hadronic spectrum above 1 GeV. 
Thanks to the ISR-method, for the first time it becomes possible 
to cover the entire mass range of interest from threshold to $4.5$ GeV in 
one single experiment. \babar~ has not only performed precision measurements
for exclusive hadronic states containing proton-antiproton, 3 pions and 4 and 6 hadrons,
but has also measured 16 $J/\psi$ and $\psi(2S)$ branching ratios, out of which
10 are better than world average. Ongoing analyses are measuring the final
states $\pi^+\pi^-$, $K^+K^-$, $\pi^+\pi^-\pi^0\pi^0$ and many more channels,
which will further improve the standard model predictions for the muon anomaly
$a_\mu$ and for the running fine structure constant $\alpha(m_Z^2)$.


\end{document}